# A Mobile Ad hoc Cloud Computing and Networking Infrastructure for Automated Video Surveillance System

**Sayed Chhattan Shah**

*Department of Information Communication Engineering,
Hankuk University of Foreign Studies, Seoul, South Korea*

**Abstract:** Mobile automated video surveillance system involves application of real-time image and video processing algorithms which require a vast quantity of computing and storage resources. To support the execution of mobile automated video surveillance system, a mobile ad hoc cloud computing and networking infrastructure is proposed in which multiple mobile devices interconnected through a mobile ad hoc network are combined to create a virtual supercomputing node. An energy efficient resource allocation scheme has also been proposed for allocation of real-time automated video surveillance tasks. To enable communication between mobile devices, a Wi-Fi Direct based mobile ad hoc cloud networking infrastructure has been developed. More specifically, a routing layer has been developed to support communication between Wi-Fi Direct devices in a group and multi-hop communication between devices across the group. The proposed system has been implemented on a group of Wi-Fi Direct enabled Samsung mobile devices.

## Introduction

Automated Video Surveillance Systems (AVSS) are used to monitor and analyze numerous situations and take necessary actions in real-time (Valera and Velastin, 2005). Compared to traditional video surveillance systems, AVSS does not require human involvement.

The AVSS comprises of object detection, object tracking, object classification, behavior analysis and action tasks. The object detection task detects the object such as a person, vehicle, or an animal in digital images and videos. Object tracking task is used to generate the trajectory of an object over time. Object classification task is used to label the detected object as a person, a group of person, vehicle, or an animal (Shah *et al.*, 2007; Teddy, 2011). One of the complicated tasks in AVSS is behavior analysis which is responsible for activity recognition and situation awareness. Based on the outcome of behavior analysis task, necessary actions are taken (Teddy, 2011; Yilmaz *et al.*, 2006). AVSS tasks are shown in Fig. 1. AVSS have applications in numerous areas including disaster management and military operations.

Due to advances in wireless communication technologies and robotics, it has become possible to use automated video surveillance system in mobile environments where multiple mobile devices such as robots and micro drones equipped with audio and video sensors are deployed to understand the situations and take necessary actions in real-time. This involves the application of computationally intensive and real-time image and video processing algorithms which require a vast quantity of computing and storage resources. To address the issue a mobile ad hoc cloud computing and networking infrastructure is proposed in which multiple mobile devices interconnected through a mobile ad hoc network are combined to create a virtual supercomputing node. To support execution of real-time automated video surveillance tasks, a resource allocation scheme has been proposed. Compared to existing schemes, the proposed scheme focuses on allocation of real time tasks and aims to reduce energy consumption.

To enable communication between mobile devices, a Wi-Fi Direct based mobile ad hoc networking infrastructure has been developed. Compared to existing mobile ad hoc networking technologies, Wi-Fi Direct provides data rates up to 250 Mbps which is sufficient for automated video surveillance systems.

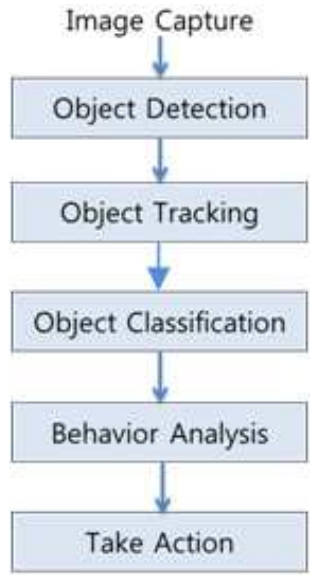

**Fig. 1:** AVSS tasks

The key contributions of this paper are as follows:

- Design of a mobile ad hoc cloud computing and a Wi-Fi Direct based mobile ad hoc cloud networking infrastructure
- An energy efficient resource allocation scheme for allocation of real-time automated video surveillance tasks
- Integration of an automated video surveillance system and a mobile ad hoc cloud system

The rest of the paper is organized as follows. Section II discusses the related work. Section III describes the mobile ad hoc cloud computing infrastructure whereas section IV focuses on mobile ad hoc cloud networking infrastructure. Integration of AVSS and mobile ad hoc cloud computing and networking infrastructure is discussed in Section V. Implementation of system is described in Section VI. Conclusion is presented in VII.

## Related Work

Numerous automated video surveillance system based on a distributed architecture has been developed (Valera and Velastin, 2005; Beynon *et al*., 2003; Lo *et al*., 2003). These systems can be classified based on numerous factors such as number of video sensors and geographic distribution of resources. Detection of events for threat evaluation and recognition (Paulidis and Morellas, 2002; Pavlidis *et al*., 2001) is a commercial system that reports abnormal behavior patterns of pedestrians and vehicles in mobile environments. It includes computer vision, threat assessment and alarms management modules. The computer vision module is responsible for the detection, tracking and classification of objects. Detection of events for threat evaluation and recognition system fuses the views of numerous cameras into one and then performs the related tasks. Crowd management with telematics imaging and communication assistance (Valera and Velastin, 2005; Cromatica, 2016) is another wide-area and multi-camera distributed video surveillance system that automatically detects dangerous situations in public transport. The system supports input from numerous devices such as CCTV, IP camera, smart sensor and audio device. A video surveillance system to automatically detect abandoned packages has been developed in (Beynon *et al*., 2003). The proposed system consists of camera view segmentation, object classification, view-object association, object tracking and abandoned packages detection modules. Compared to existing systems, our proposed system can be deployed in mobile ad hoc environments.

In literature, several resource allocation schemes has been developed for allocation of tasks to mobile ad hoc systems. Authors in (Rodriguez *et al*., 2012) have proposed several job stealing techniques to reduce the processing energy consumption. Authors in (Hariharasudhan *et al*., 2015) have proposed a scheme that addresses network connectivity, node mobility and energy consumption problems. To deal with uncertainty, an idea of application waypoints has been introduced in (Ghasemi-Falavarjani *et al*., 2015). A service provider node executing application task reports to a broker node with an estimate of residual task completion time. If broker does not receive feedback about the estimated residual task completion time from the service provider node at the specified waypoint, it marks service provider as failed and assigns additional resources to take over the incomplete tasks. In (Shah and Park, 2011) authors have proposed a resource allocation scheme that aims to reduce communication energy consumption. The scheme uses dynamic transmission power and hybrid architecture for effective decisions. The node mobility problem is addressed in (Shah *et al*., 2012). The scheme proposed in (Shah *et al*., 2012) is divided into two phases. The first phase exploits the history of user mobility patterns to select nodes that will remain connected for long time and the second phase considers application tasks and distance between nodes to reduce communication cost. An efficient and robust resource allocation scheme to address node mobility and energy consumption problems has been developed in Rodriguez *et al*. (1999). Compared to existing schemes, our proposed scheme focuses on allocation of real-time tasks and aims to reduce energy consumption.

Existing mobile ad hoc networking technologies such as ZigBee and IEEE 802.11b provide limited bandwidth and therefore are not suitable for mobile automated video surveillance system applications. Wi-Fi Direct is a new device-to-device technology that aims provides data rates up to 250 Mbps (WFP2P, 2014). Wi-Fi Direct technology however does not provide support for multi-hop communication and group client-to-group client communication.

To overcome the limitations of Wi-Fi Direct, several systems (Duan *et al.*, 2014; Jung *et al.*, 2014; Casetti *et al.*, 2015; Funai *et al.*, 2015; Felice *et al.*, 2016) have been proposed. Authors in (Duan *et al.*, 2014) have developed a content centric ad hoc network that includes mobile devices equipped with Wi-Fi Direct technology. It is claimed that a group client connected to a group owner on a channel x can communicate to another group client on channel y using a concurrent mode (WFD, 2017). The proposed system is implemented on network simulator ns-3 which provides support for Wi-Fi ad hoc mode (Duan *et al.*, 2014) only. Another Wi-Fi Direct based mobile ad hoc network architecture is proposed in (Jung *et al.*, 2014). The proposed architecture uses a tunneling mechanism to allow inter-group communication between Android smartphones. To support multi-hop communication, a simple version of Destination-Sequenced Distance-Vector Routing protocol (Mohapatra and Kanungo, 2012) has been implemented. A multi-group networking scheme has been proposed in (Casetti *et al.*, 2015). The scheme uses CCN module to send and receive contents. CCN module consists of Content Routing Table (CRT) and Pending Interest Table (PIT). CRT stores IP addresses of nodes in a communication range. PIT records information to route content to requester by storing IP address of a node from which a request was received. The schemes proposed in (Jung *et al.*, 2014) and (Casetti *et al.*, 2015) does not provide support for communication across the group.

## Mobile Ad hoc Cloud Computing Infrastructure

A distributed system consists of a collection of autonomous computers, connected through a network and distribution middleware, which enables computers to coordinate their activities and to share the resources of the system, so that users perceive the system as a single, integrated computing facility (Paulidis and Morellas, 2002). The distributed systems are divided into three main categories: cluster, grid and cloud. In cluster, distributed computing devices are connected through a local area network whereas in grid, geographically distributed resources are connected through a wide area network. The cloud computing has evolved from cluster and grid computing and is integration of various concepts and technologies such as hardware virtualization, utility computing, autonomic computing, pervasive computing and service-oriented architecture. In cloud computing, everything from computing power to communication infrastructure and applications are delivered as a service over a network.

Cloud computing systems have been used to solve large and complex problems. These systems include powerful computing resources connected through high speed networks. Due to recent advances in mobile computing and networking technologies, it has become feasible to integrate various mobile devices such as robots, aerial vehicles, sensors and smart phones with cloud computing systems. The approaches for integrating mobile devices with cloud computing systems are divided into two main categories: mobile cloud computing (Hariharasudhan *et al.*, 2015; Shah, 2017) and mobile ad hoc cloud computing (Ghasemi-Falavarjani *et al.*, 2015; Shah and Park, 2011; Shah, 2017).

In mobile cloud computing, mobile devices are integrated with a cloud computing system through an infrastructure-based communication network such as cellular network. The integration of mobile devices with cloud computing systems enable mobile devices to access vast amount of processing power and storage space. This makes possible to execute data and computationally intensive applications such as image and video processing on mobile devices. Data storage and execution of such applications on cloud also improve reliability and extend the battery life of mobile devices. The system architecture of mobile cloud computing system is given in Fig. 2.

Mobile nodes such as robot or smart phone are connected to an internet cloud via a pre-existing network infrastructure-based system such as Wi-Fi access point, base station or satellite.

The mobile cloud computing systems are restricted to infrastructure-based communication systems and therefore cannot be used in mobile ad hoc environments.

In mobile ad hoc cloud computing (Shah, 2013), multiple mobile devices interconnected through a mobile ad hoc network are combined to create a virtual supercomputing node. The system architecture of mobile ad hoc cloud computing system is given in Fig. 3.

The layered architecture of mobile ad hoc cloud computing system is presented in Fig. 4. Mobile nodes communicate with each other through a mobile ad hoc network which provides several communication and networking services such as network discovery, monitoring and routing. The cloud middleware layer is responsible for resource management, failure management, mobility management, communication management and task migration. In addition, it hides all the complexities and provides a single system image to user and applications running on the system.

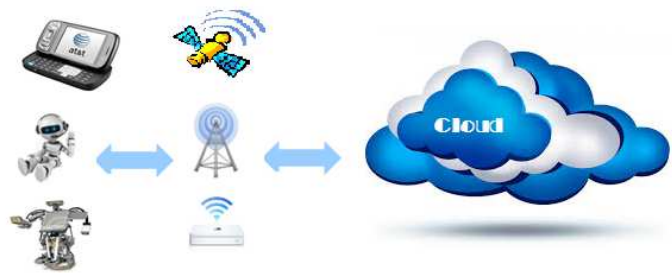

**Fig. 2:** Mobile cloud computing system architecture

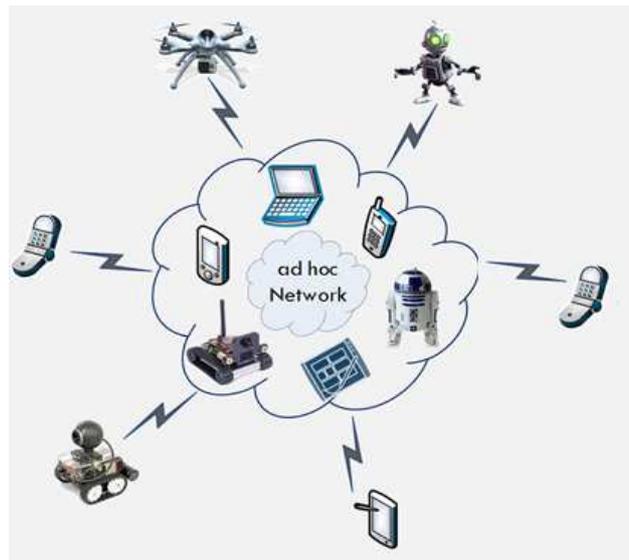

**Fig. 3:** Illustrates a diagram of mobile ad hoc cloud in which multiple mobile devices interconnected through a mobile ad hoc network are presented as a powerful, unified computing resource

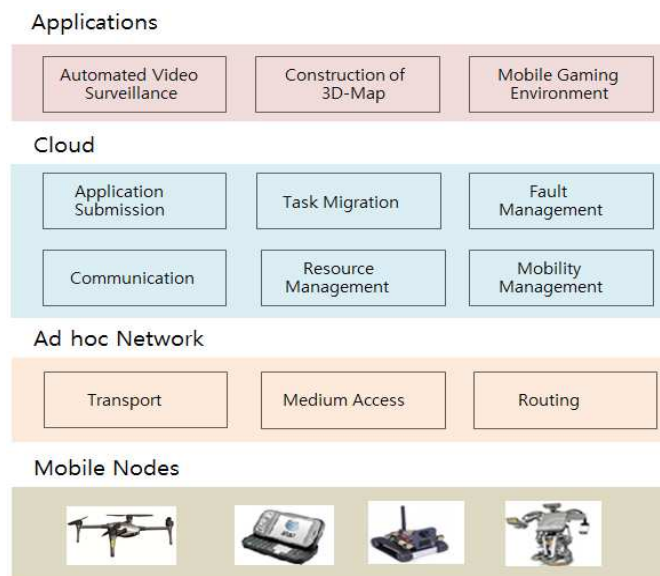

**Fig. 4:** Mobile ad hoc cloud computing system architecture

770

# Wi-fi Direct based Mobile Ad hoc Cloud Networking Infrastructure

A mobile ad hoc network is a wireless network of mobile devices that communicate with each other without any pre-existing network infrastructure. Existing mobile ad hoc networking technologies such as ZigBee and IEEE 802.11b provide limited bandwidth and therefore are not suitable for mobile automated video surveillance system applications (Shah, 2016). Wi-Fi Direct is a new device-to-device technology that provides data rates up to 250 Mbps (WFP2P, 2014). The characteristics of existing wireless communication standards and Wi-Fi Direct technology are given in Table 1.

This section provides an overview of Wi-Fi Direct technology and proposes a Wi-Fi Direct based mobile ad hoc cloud networking infrastructure.

*Overview of WI-FI Direct Technology*

Wi-Fi Direct is a new technology standardized by Wi-Fi Alliance (Shah, 2016; Conti *et al.*, 2013; Camps-Mur *et al.*, 2013). It enables mobile devices to directly communicate with each other without a wireless access point.

In order to communicate, Wi-Fi Direct devices create a group and negotiate a role. One device in a group act as a Group Owner (GO) while remaining devices play a role of Group Client (GC). A new node can join a group anytime as a client. Group owner can simultaneously communicate with multiple group clients and a group client can communicate with a group owner. The communication patterns between devices in a group are depicted in Fig. 5.

Wi-Fi Direct devices can also use a concurrent mode to communicate with each other via multiple wireless communication technologies. Wi-Fi Direct concurrent mode is depicted in Fig. 6.

*Device Discovery Process*

The purpose of device discovery process is to discover devices in a communication range. The discovery process consists of scan phase and find phase.

During a scan phase, a device scans wireless channels to discover information about nearby devices. The scan phase also supports discovery of legacy devices operating on channels in addition to social channels (Camps-Mur *et al.*, 2013).

During a find phase, a device alternates between search state and listen state. In search state, each device sends a probe request message. In listen state, a device listens to probe request messages and replies with a probe response message. The time for each state is randomly distributed between 100 and 300ms (Camps-Mur *et al.*, 2013).

**Table 1:** Characteristics of wireless communication standards

|  | IEEE 802.11a | IEEE 802.11b | Wi-Fi direct |
|---|---|---|---|
| **Max range** | 45m | 45m | 200m |
| **Max data rate** | 54Mbps | 11Mbps | 250Mbps |

*Group Formation Process*

Wi-Fi Direct technology support three procedures for a group formation: Standard, autonomous and persistent. In a standard procedure, a device discovery process is followed by a group owner role negotiation process. In autonomous group formation procedure, a node elects itself as a group owner and then announces its presence through beacon messages. In persistent procedure, a device declares a group as persistent using an attribute present in a beacon frame. This paper focuses on the standard group formation procedure (Conti *et al.*, 2013).

In standard group formation procedure, a group owner role negotiation process involves three-way handshake: Request-Response-Confirmation. The devices decide the role of group owner and the channels on which group will operate. To decide group owner role, devices share a group owner *intent value* and the device with maximum intent value take the role.

Once group owner role negotiation process is complete, an authentication procedure is performed to establish a secure wireless connection (Duan *et al.*, 2014). This is followed by an address configuration phase in which devices receive an IP address assigned by DHCP server running on a group owner (Camps-Mur *et al.*, 2013).

*Wi-Fi Direct Technology Limitations*

Wi-Fi Direct technology does not provide support for communication between group clients. Group clients can only communicate with a group owner as shown in Fig. 7a.

Wi-Fi Direct technology also does not provide support for multi-hop communication. As shown in Fig. 7b, it is possible for a group owner GO to communicate with a device D via a group client GC but it is not supported.

*Wi-Fi Direct based Mobile Ad hoc Cloud Networking Infrastructure*

To overcome the limitations of Wi-Fi Direct technology, a Wi-Fi Direct based mobile ad hoc cloud network infrastructure has been developed. More specifically, a routing layer has been developed to support communication between Wi-Fi Direct devices in a group and multi-hop communication between devices across the group.

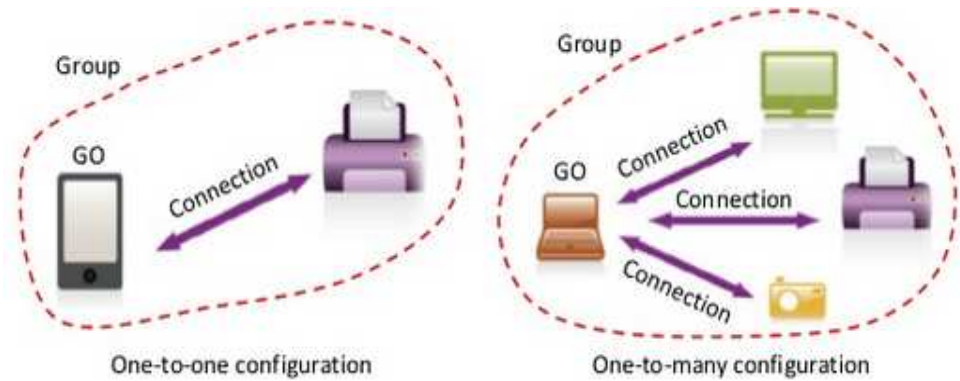

**Fig. 5:** Communication in a group

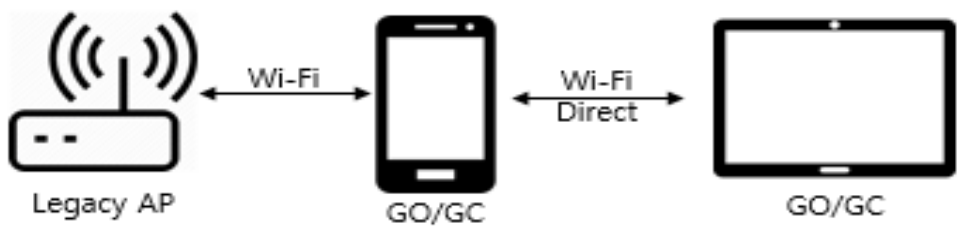

**Fig. 6:** Wi-Fi direct concurrent mode

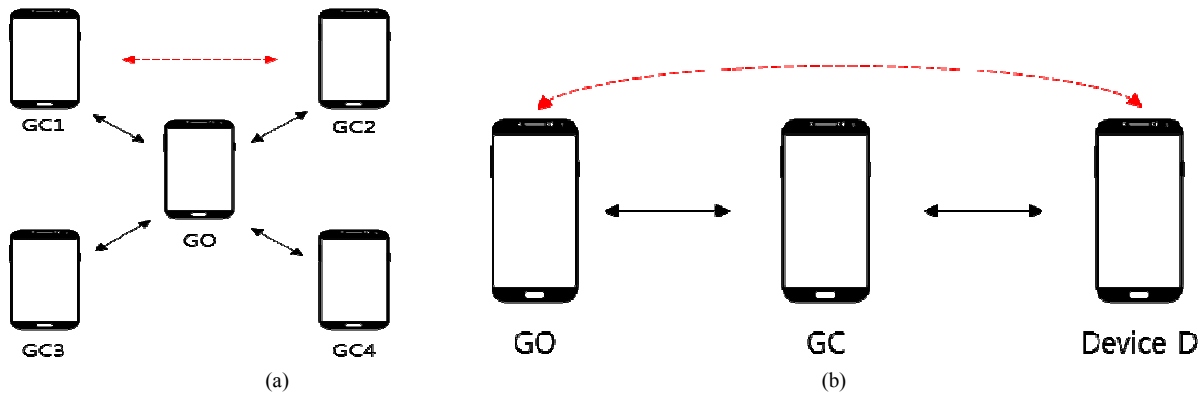

**Fig. 7:** Wi-Fi direct technology limitations, (a) D2D communications between GO and GCs (b) Multi-hop communications

The architecture of a routing layer is given in Fig. 8. It consists of discovery manager, routing manager, routing table, data transfer manager, application data manager and connection manager. The routing layer accesses the Wi-Fi Direct technology and provides communication and networking services to an application layer. The architectural elements of routing layer are described below whereas relationship between architectural elements is depicted in Fig. 9.

*Discovery Manager*

Discovery manager discovers devices in the network. To discover devices, discovery manager periodically broadcast a discovery request packet which includes source address, destination address and sequence number. A node receiving a discovery request packet replies with a discovery reply packet. The discovery reply packet includes source address, destination address, sequence number, a list of neighbor nodes and distance in hops. A node receiving a reply packet add a new entry or updates an existing entry in the routing table. To reduce communication cost only updated or new information stored in routing table is included in a discovery reply packet. The discovery manager also discovers information about communication links between devices which is then used to estimate available bandwidth and transmission energy consumption.

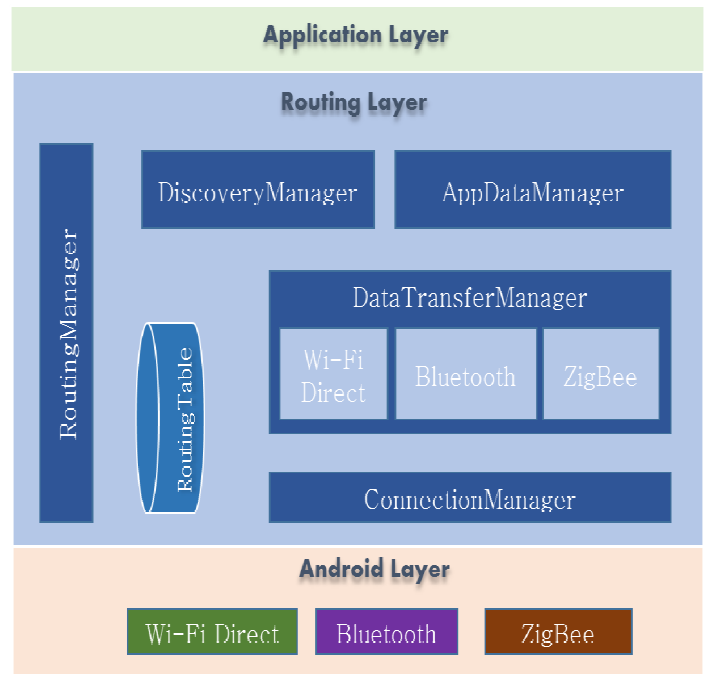

**Fig. 8:** Proposed system architecture

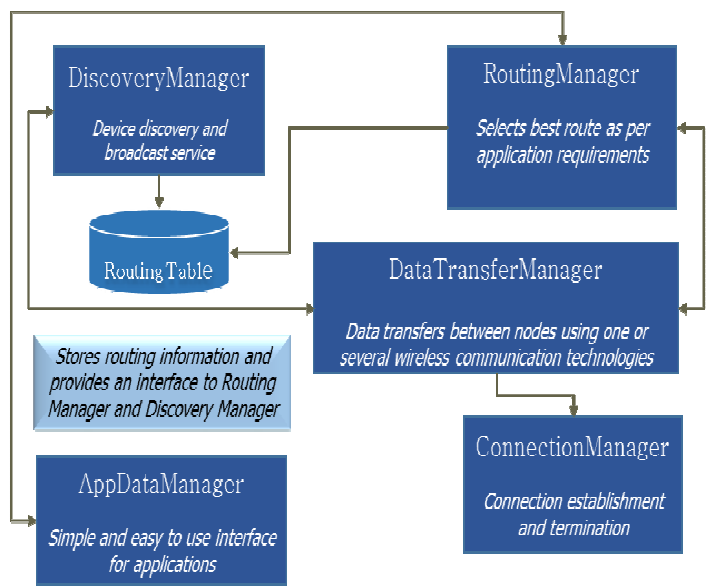

**Fig. 9:** Relationship between routing layer components

*Routing Manager*

Routing manager implements a routing algorithm and is responsible for the selection of a route to a destination. The discovery manager stores discovered information in a routing table whereas routing manager uses routing table information to make effective routing decisions. For real-time tasks, a route with a minimum latency is selected whereas for non-real-time tasks, a route with several hops is selected to reduce transmission energy consumption and increase node life time.

*Routing Table*

The information stored in a routing table is given below.

| Next node | Destination node | Sequence No | Hops | Available bandwidth |
| --- | --- | --- | --- | --- |

*Data Transfer Manager*

Data transfer manager is responsible for the transmission of data across the network. It is used by a routing manager and discovery manager to transmit packets to nearby nodes. The data transfer manager communicates with Wi-Fi Direct technology for data transmission and provides a simple and easy to use interface to routing layer elements. It can be easily extended to support communication between devices using various communication technologies such as ZigBee and Bluetooth as shown in Fig. 10.

*Application Data Manager*

Application data manager is an interface between application layer and routing layer. Application sends data to application data manager for transmission across the network. The application data manager then communicates with routing manager for data transmission.

*Connection Manager*

Connection manager is responsible for creation, maintenance and termination of a network connection.

# Integration of AVSS and Mobile ad hoc Cloud System

*Automated Video Surveillance System Model*

Automated video surveillance system consists of object detection, object tracking, object classification, video indexing, behavior analysis and action tasks. The execution of these tasks require a vast quantity of computing and storage resources. To address the issue, a mobile ad hoc cloud computing and networking infrastructure is developed.

To support the execution of automated video surveillance system on a mobile ad hoc cloud, automated video surveillance system application is divided into parallel and interdependent tasks as shown in Fig. 11.

For object detection task, adaptive background subtraction (Stauffer and Grimson, 1999) algorithm has been used. SIFT algorithm (Lowe, 1999) has been used for object tracking task while algorithms in (Dedeoğlu, 2006) have been adopted for real-time object classification and human action recognition tasks. The detailed view of AVSS tasks is provided in Fig. 12.

*Mobile Ad hoc Cloud System Model*

Mobile ad hoc cloud includes numerous mobile devices such as a mobile robot, micro aerial vehicle and smartphone. Mobile devices are equipped with a Wi-Fi Direct technology and are heterogeneous in terms of computational power, storage space and battery power. Nodes communicate with each other through a Wi-Fi Direct based mobile ad hoc cloud network. The communication is achieved by passing messages and communication between tasks assigned to the same node is negligible.

*Resource Allocation Scheme*

The objective of resource allocation scheme is to select a node for execution of a task. Compared to existing schemes, the proposed scheme considers allocation of real-time tasks and aims to reduce energy consumption. The resource allocation process consists of two steps: (1) selection of nodes on which estimated task execution time is less than the task deadline and (2) given the list of nodes selected in step 1, allocate a task to a node on which estimated energy consumption is minimum. To estimate task execution time and energy consumption, models proposed in (Ghasemi-Falavarjani *et al.*, 2015) have been adopted.

*Notations*

| | |
|---|---|
| $|N|$ | Set of mobile nodes |
| $|T|$ | Set of tasks |
| $t_i$ | A task within an application |
| $n_i$ | A node |
| $P_{Node_i}$ | Processing power of node $n_i$ |
| $T_{Size_i}$ | Size of a task $t_i$ |
| $E_{Node_i}$ | Available energy at |
| $E_{threshold}$ | Threshold value for energy |
| $DiS_{t_i}$ | Input data size of task $t_i$ |
| $DoS_{t_i}$ | Output data size of task $t_i$ |
| $D_{t_i}$ | Deadline of task $t_i$ |
| $a_i$ | Processing energy consumption per time unit |
| $b_i$ | Transmission energy consumption per packet |
| $DTT_{t_i}$ | Data transfer time for task $t_i$ |
| $EE(t_i, n_i)$ | Estimated execution time of task $t_i$ on node |
| $EEC(t_i,n_i)$ | Estimated energy consumption of task $t_i$ on node |

*Cost Estimation Models*

Models proposed in (Ghasemi-Falavarjani *et al.*, 2015) have been adopted to estimate task execution time and energy consumption:

$$EE(t_i, n_i) = \frac{T_{Size_i}}{P_{Node_i}} + DTT_{t_i} \qquad (1)$$

$$DTT_{t_i} = \frac{DiS_{t_i}}{B_{channel}} + \frac{DoS_{t_i}}{B_{channel}} \qquad (2)$$

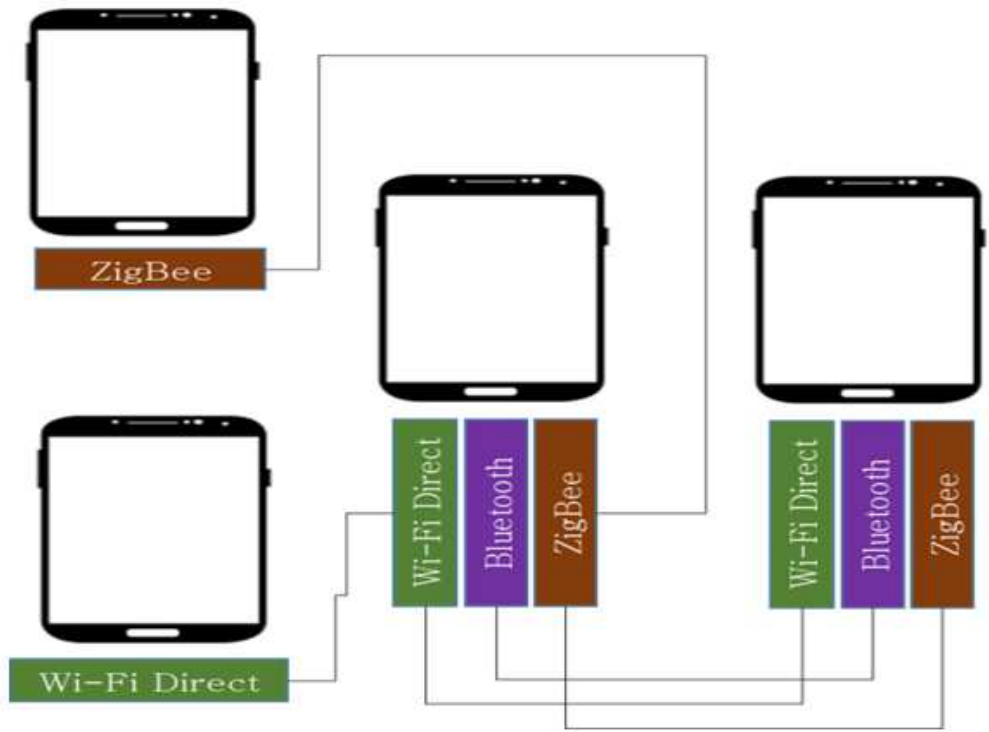

**Fig. 10:** An illustration of heterogeneous network environment

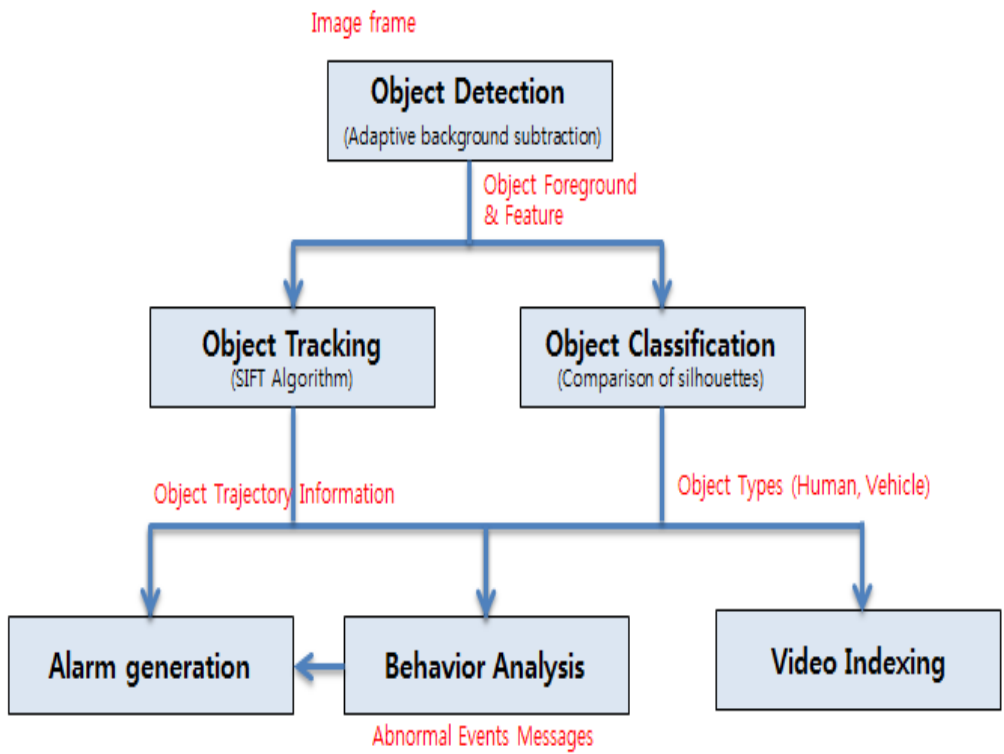

**Fig. 11:** AVSS for mobile ad hoc cloud

775

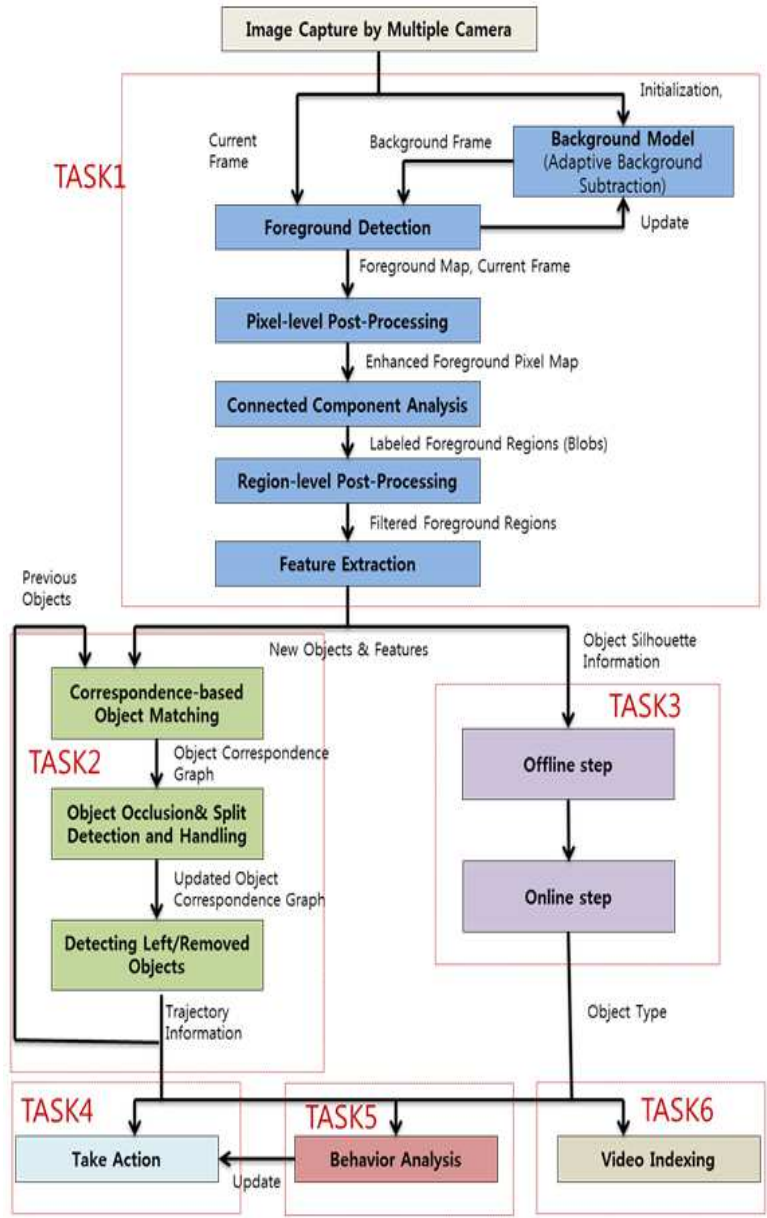

**Fig. 12:** Detailed view of AVSS tasks

$$EEC(t_i, n_i) = a_i * \frac{T_{Size_i}}{P_{Node_i}} + b_i * Pkt_{num} \quad (3)$$

$$Pk_{num} = \frac{DiS_{t_i}}{Pkt_{size}} + \frac{DoS_{t_i}}{Pkt_{size}}$$

*Resource Allocation Algorithm*

**Input:** Set of mobile nodes |**N**|
 Set of tasks |**T**|
**Output:** Task $t_i$ assigned to node
Sort tasks in a set |**T**|
**While** (|**T**| is not empty) **do**
 Get a task $t_i$ from task set |**T**|
 **For** each node $n_i \in$|**N**|
  Estimate execution time of task $t_i$ on node $n_i \in$|**N**| using equation 1
  Estimate energy consumption of task $t_i$ on node $n_i \in$ |**N**| using equation 3
  **If** (Estimated execution time of task $t_i$ on node $n_i <= D_{t_i}$) and (Estimated energy consumption of task $t_i$ on node < Estimated energy consumption of task $t_i$ on node $n_{preSel}$)
   Allocate to task $t_i$ to node $n_i$

## Implementation

Automated video surveillance system comprising of object detection, object tracking and object classification tasks have been implemented on a single node and on a mobile ad hoc cloud system consisting of three mobile nodes. The characteristics of mobile nodes are given in Table 2 whereas task-node assignments are shown in Fig. 15. An adaptive background subtraction algorithm proposed in (Stauffer and Grimson, 1999) has been implemented for object detection task and SIFT algorithm developed in (Lowe, 1999) has been implemented for object tracking task. For real-time object classification and human action recognition tasks, algorithms developed in (Dedeoğlu, 2006) have been adopted. Mobile devices are equipped with a Wi-Fi Direct technology and are heterogeneous in terms of computational power, storage space and battery power.

In order to enable direct and multi-hop communication between devices, the discovery manager, the routing manager, the routing table, the data transfer manager and the application data manager has been implemented. A data transfer manager provides an interface to Wi-Fi Direct technology and implements send, receive and broadcast functions for transmission of data across the network. The information stored in the routing table is given below.

| Next node | Destination node | Sequence No | Hops |
|---|---|---|---|

Mobile ad hoc cloud network formation begins with a discovery process. To discover devices, each node periodically broadcasts a discovery request packet which includes source address, destination address and sequence number. A node receiving a discovery request packet replies with a discovery reply packet. The discovery reply packet includes source address, destination address, sequence number, a list of neighbor nodes and distance in hops. A node receiving a reply packet add a new entry or updates an existing entry in the routing table.

To verify multi hop communication, a simple data transfer application has been developed. The application running on a source node sent a data transmission request to application data manager. The data transmission request consisted of data and destination address. The application data manager communicated with the routing manger which selected a next node based on information stored in the routing table. Packet processing process involving routing layer components is shown in Fig. 13.

## Results

The execution time of automated video surveillance system tasks on a single node and on a mobile ad hoc cloud system is given in Fig. 14.

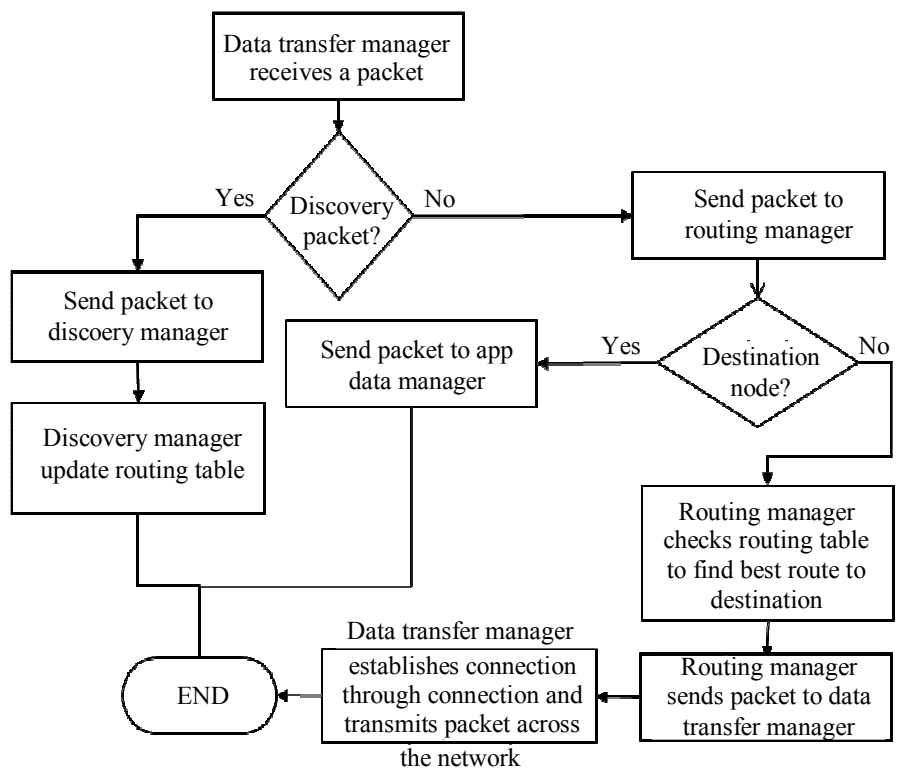

**Fig. 13:** Packet processing diagram

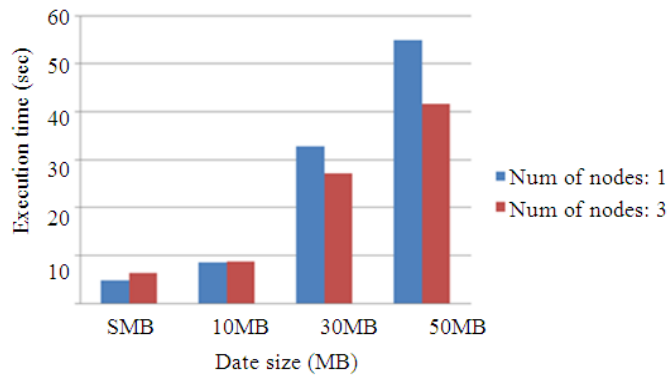

**Fig. 14:** Execution time of AVSS

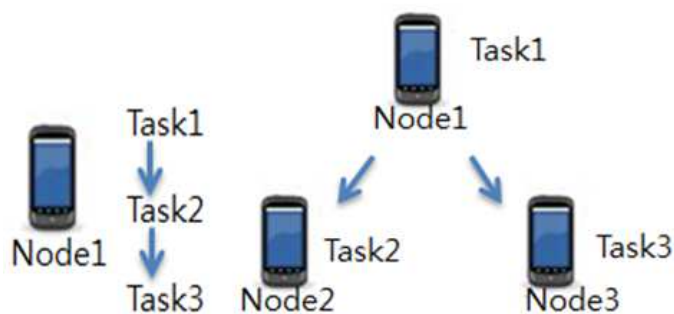

**Fig. 15:** Task-node assignments

**Table 2:** Characteristics of mobile nodes

| Devie | OS | CPU | RAM |
| --- | --- | --- | --- |
| Galaxy S7 | Android v7.0 | Octa (2.4GHz, 1.6GHz) | 4GB |
| Galaxy S2 | Android v4.1 | Dual (1.2GHz) | 1GB |
| Vega LTE | Android v4.1 | Dual (1.5GHz) | 1GB |

As results show, automated video surveillance system on mobile ad hoc cloud significantly reduces execution time and improves performance by 17% with 30 megabytes of data and 20% with 50 megabytes of data.

The purpose of the experiments was to check the feasibility of proposed system. In future work, we aim to implement all the tasks of automated video surveillance system and conduct extensive experiments in wide range of application and network scenarios.

## Conclusion

In this study, a mobile cloud computing and networking infrastructure has been proposed for automated video surveillance system. To support execution of real-time automated video surveillance tasks, a resource allocation scheme has also been proposed. Compared to existing schemes, the proposed scheme focuses on allocation of real time tasks and aims to reduce energy consumption. To enable communication between mobile devices, a Wi-Fi Direct based mobile ad hoc cloud networking infrastructure has been developed. More specifically, a routing layer has been developed to support communication between Wi-Fi Direct devices in a group and multi-hop communication between devices across a group.

The routing layer consists of discovery manager, routing manager, routing table, data transfer manager, application data manager and connection manager. The routing layer accesses the Wi-Fi Direct technology and provides communication and networking services to an application layer. A simple automated video surveillance system comprising of object detection, object tracking and object classification tasks have been implemented on a single node and on a mobile ad hoc cloud system consisting of three mobile nodes.

## Acknowledgment

This work was supported by Hankuk University of Foreign Studies Research Fund of 2017 and National Research Foundation of Korea (2017R1C1B5017629).

Sayed Chhattan Shah / Journal of Computer Sciences 2017, 13 (12): 767.780
DOI: 10.3844/jcssp.2017.767.780
This paper is an extended version of paper "A mobile ad hoc cloud for automated video surveillance system" published in proceedings of *2017 International Conference on Computing, Networking and Communications (ICNC)*, Santa Clara, CA, 2017, pp. 1001-1005.

## Ethics

There are no ethical issues in publishing and giving open Access.

## References

Beynon, M.D., D.J. Van Hook, M. Seibert, A. Peacock and D. Dudgeon, 2003. Detecting abandoned packages in a multi-camera video surveillance system. Proceedings of the IEEE Conference on Advanced Video and Signal Based Surveillance, Jul. 22-22, IEEE Xplore Press, Miami. DOI: 10.1109/AVSS.2003.1217925

Camps-Mur, D., A. Garcia-Saavedra and P. Serrano, 2013. Device-to-device communications with wi-fi direct: Overview and experimentation. IEEE Wirel. Commun., 20: 96-104.
DOI: 10.1109/MWC.2013.6549288

Casetti, C., C.F. Chiasserini, L.C. Pelle, C.D. Valle and Y. Duan *et al.*, 2015. Content-centric routing in wi-fi direct multi-group networks. Proceesings of the IEEE 16th International Symposium on Mobile and Multimedia Networks, (MMN' 15), pp: 1-9.

Conti, M., F. Delmastro, G. Minutiello and R. Paris, 2013. Experimenting opportunistic networks with wi-fi direct. Proceedings of the IFIP Wireless Days, Nov. 13-15, IEEE Xplore Press, Valencia, Spain.
DOI: 10.1109/WD.2013.6686501

Cromatica, http://cordis.europa.eu/result/rcn/25106_en.html, accessed in July 2016.

Dedeoğlu, Y., 2006. Silhouette-based method for object classification and human action recognition in video. Proceedings of the ECCV 2006 Workshop on HCI, Graz, Austria, May 13-13.

Duan, Y., C. Borgiattino, C. Casetti, C.F. Chiasserini and P. Giaccone *et al.*, 2014. Wi-fi direct multi-group data dissemination for public safety. Proceedings of the World Telecommunications Congress, Jun. 1-3, IEEE Xplore Press, Berlin.

Felice, M.D., L. Bedogni and L. Bononi, 2016. The emergency direct mobile App: Safety message dissemination over a multi-group network of smartphones using wi-fi direct. Proceesings of the 14th ACM International Symposium on Mobility Management and Wireless Access, Nov. 13-17, IEEE Xplore Press, Malta, pp: 99-106.
DOI: 10.1145/2989250.2989257

Funai, C., C. Tapparello and W. Heinzelman, 2015. Supporting multi-hop device-to-device networks through wifi direct multi-group networking. University of Rochester.

Ghasemi-Falavarjani, S., M. Nematbakhsh and B.S. Ghahfarokhi, 2015. Context-aware multi-objective resource allocation in mobile cloud. Comput. Electrical Eng., 44: 218-240.
DOI: 10.1016/j.compeleceng.2015.02.006

Hariharasudhan, V., E.K. Lee, I. Rodero and D. Pompili, 2015. Uncertainty-aware autonomic resource provisioning for mobile cloud computing. IEEE Trans. Parallel Distributed Syst., 26: 2363-2372.
DOI: 10.1109/TPDS.2014.2345057

Jung, W.S., H. Ahn and Y.B. KO, 2014. Designing content-centric multi-hop networking over wi-fi direct on smartphones. Proceedings of the IEEE Conference Wireless Communications and Networking, Apri . 6-9, IEEE Xplore Press, Istanbul, Turkey.
DOI: 10.1109/WCNC.2014.6952920

Lo, P.L.B., J. Sun and S.A. Velastin, 2003. Fusing visual and audio information in a distributed intelligent surveillance system for public transport systems. Acta Automatica Sinica, 29: 393-407.

Lowe, D.G., 1999. Object recognition from local scale-invariant features. Proceedings of the 7th IEEE International Conference on Computer Vision, Sept. 20-27, IEEE Xplore Press, Kerkyra, Greece, pp: 1150-1157. DOI: 10.1109/ICCV.1999.790410

Mohapatra, S. and P. Kanungo, 2012. Performance analysis of AODV, DSR, OLSR and DSDV Routing Protocols using NS2 Simulator, Procedia Eng., 30: 69-76. DOI: 10.1016/j.proeng.2012.01.835

Paulidis, I. and V. Morellas, 2002. Two examples of indoor and outdoor surveillance systems: Motivation, design and testing. Proceedings of the Video-based Surveillance Systems, Kluwer Academic Publishers, (KAP' 02), Boston.

Pavlidis, I., V. Morellas, P. Tsiamyrtzis and S. Harp, 2001. Urban surveillance systems: From the laboratory to the commercial world. Proc. IEEE, 89: 1478-1497. DOI: 10.1109/5.959342

Rodriguez, J.M., C. Mateos and A. Zunino, 1999. Energy-efficient job stealing for cpu-intensive processing in mobile devices. Computing, 96: 87-117.
DOI: 10.1007/s00607-012-0245-5

Rodriguez, J.M., C. Mateos and A. Zunino, 2012. Energy-efficient job stealing for CPU-intensive processing in mobile devices. Computing, 96: 87-117.
DOI: 10.1007/s00607-012-0245-5

Shah, M., O. Javed and K. Shafique, 2007. Automated visual surveillance in realistic scenarios. IEEE Comput. Society.
779